\documentstyle[sprocl]{article} \input{amssym.def} \input{amssym.tex}

\renewcommand{\ll}{\label} \newcommand{\C}{\Bbb C}
\newcommand{\R}{\Bbb R} \newcommand{\inv}{^{-1}}
\newcommand{\be}{\begin{equation}} \newcommand{\ee}{\end{equation}}
\newcommand{\bea}{\begin{eqnarray}} \newcommand{\eea}{\end{eqnarray}}
 \newcommand{\bib}{\bibitem}
\newcommand{\ci}{\cite} 
\newcommand{\ca}{$C^*$-algebra} \newcommand{\rep}{representation}

\newcommand{\Hs}{Hilbert space} \newcommand{\mom}{momentum map}
 
 \newcommand{\til}{\tilde}
\newcommand{\raw}{\rightarrow}

\newcommand{\ot}{\otimes} 
\newcommand{\la}{\langle} \newcommand{\ra}{\rangle}
\newcommand{\x}{\times} 
 
\newcommand{\cci}{C^{\infty}_c} \newcommand{\half}{\mbox{\footnotesize
$\frac{1}{2}$}}  
\newcommand{\al}{\alpha} 
 
 \newcommand{\Dl}{\Delta}

 \newcommand{\kp}{\kappa}
\newcommand{\lm}{\lambda} 
 \newcommand{\sg}{\sigma}
  
\newcommand{\Ph}{\Phi} \newcommand{\phv}{\varphi}
  \newcommand{\Ps}{\Psi}

\newcommand{\CE}{{\cal E}} \newcommand{\CN}{{\cal N}}
\newcommand{\CV}{{\cal V}} \renewcommand{\H}{{\cal H}}
\begin{document}
\title{QUANTIZATION OF SINGULAR SYSTEMS AND INCOMPLETE MOTIONS}
\author{N.P. LANDSMAN} \address{Korteweg-de Vries Institute for
Mathematics, University of Amsterdam, Plantage Muidergracht 24, 1018
TV AMSTERDAM, THE NETHERLANDS \\E-mail: npl@wins.uva.nl}
\maketitle\abstracts{ The need for a mathematically rigorous
quantization procedure of singular spaces and incomplete motions is
pointed out in connection with quantum cosmology.  We put our previous
suggestion for such a procedure, based on the theory of induced \rep s
of \ca s, in the light of L. Schwartz' theory of Hilbert subspaces.
This turns out to account for the freedom in the induction procedure,
at the same time providing a basis for generalized eigenfunction
expansions pertinent to the needs of quantum cosmology. Reinforcing
our previous proposal for the wave-function of the Universe, we are
now able to add a concrete prescription for its calculation.  }
\section{Introduction}
\subsection{Incomplete dynamical systems}
 Minisuperspace cosmologies with singularities are examples of
dynamical systems whose motion may be incomplete (that is, {\em in a given
parametrization} the flow may not be defined for all $t\in\R$).  In
general, classical incomplete motion may occur even on smooth phase
spaces; for example, a particle may escape to infinity in a finite
time.  In cosmological examples, though, the incompleteness is
generically caused by the presence of singularities in the space of
physical degrees of freedom of the model in question. The notion of
singularity used here is rather wide; boundary points to a manifold
are included as possible singular points. Indeed, the incompleteness
of the motion of a cricket ball off a batsman hitting a boundary is
precisely caused by the boundary of the field. On the other hand, one
may think of the incompleteness of the motion of an observer falling
into a black hole.  The question whether the space is question may be
extended so as to render the given motion complete is interesting, but
not quite relevant in this context.

In the general theory of dynamical systems, incompleteness is not much
studied, for the simple reason that under mild conditions one may
reparametrize the flow, so as to make the motion complete.  The
question whether a singularity is approached in finite time therefore
appears to be ill-defined.  However, the physics of the situation
usually singles out a preferred choice of the time co-ordinate, an
issue that has been much discussed in classical and quantum gravity;
see Isham \ci{Isham} and references therein.

In our opinion, in a canonical context the choice of a time parameter
is determined by the choice of the physical Hamiltonian, which simply
defines time through Hamilton's equations.  Hence the above question
in any case makes sense in the setting of {\em Hamiltonian} dynamical
systems. Further to the well-studied classes of integrable and chaotic
systems, one should therefore study the class of incomplete
Hamiltonian dynamical systems. While this is fascinating already
classically, the issue of genuine physical importance lies in the
quantum theory of such systems.  It is well known that, in view of the
unitarity of the time-evolution, a quantum system can be neither
chaotic (in the classical sense) nor incomplete. 
 Hence one faces the difficult task of
identifying properties of the quantum Hamiltonian which indicate that
the classical limit of the theory is integrable, chaotic, or
incomplete. Signatures for chaos are found in the statistical
distribution of the eigenvalues, whereas classical incompleteness is
often related to the fact that the Hamiltonian fails to be essentially
self-adjoint on its natural domain of smooth compactly supported
wave-functions \ci{ReedSimon75}.  It is intuitively evident why this
is so: such wave-functions have support away from the singularities,
so that the natural domain contains no information about the boundary
conditions \ci{Berezanski}.
\subsection{Singular reduction}
Our second source of inspiration lies in the origin of the
singularities of the physical phase space of general
relativity. Namely, since the space of unphysical degrees of freedom
(in whatever formalism one uses) is regular, the singularities come
from the constraints of Einstein gravity.  The singularities of the
reduced phase space are well understood: the physical phase space is
stratified by symplectic manifolds, each of which is stable under the
Hamiltonian flow generated by any function that Poisson-commutes with
all constraints \ci{IsenbergMarsden}.  This important
infinite-dimensional fact is entirely analogous to the case of
singular Marsden-Weinstein quotients by proper actions of
finite-dimensional Lie groups on finite-dimensional symplectic
manifolds \ci{Cushman}.

When the symplectic space one starts from is a cotangent bundle
$S=T^*Q$, and the group action on $S$ (with equivariant momentum map $J$)
is pulled back 
from a $G$-action on $Q$, the Marsden-Weinstein quotient at zero is
simply $J\inv(0)/G=T^*(Q/G)$, where away from the singular points the
cotangent bundle is defined as usual, whereas at the singular points
it is defined by the left-hand side.  One may quantize the reduced
space by saying that the pertinent Hilbert space is $L^2(Q/G)$, and
that the Hamiltonian is minus the Laplacian, defined on the domain
$\cci(Q/G)$ \ci{EmmrichRomer}.  As already mentioned in the previous
subsection, this operator is not essentially self-adjoint; one needs
to choose boundary conditions, so there is no unique quantum
theory.

Instead, one needs a method of constrained quantization which
parallels the classical procedure of symplectic reduction, and which
still applies when the usual conditions guaranteeing that the reduced
space be smooth \ci{Cushman} do not apply.  The method proposed by the
author \ci{JGP,MT,WW} satisfies these requirements, and simultaneously
solves the problem described in the previous paragraph.

The essence of this method is best explained by comparing it with
Dirac's well-known technique of constrained quantization \ci{Dirac}.
We restrict ourselves to first-class constraints.  Recall that
symplectic reduction is a two-step procedure: firstly, the constraints
$\phv_i=0$ are imposed, and secondly, roughly speaking,
gauge-equivalent points on the constraint hypersurface are identified.
In quantum theory only one of these steps has to be taken. Dirac's
approach to constrained quantization selects the first step, in
imposing the quantized constraints $\hat{\phv}_i$ as state conditions
$\hat{\phv}_i\Ps=0$ on the \Hs\ $\H$ of the unconstrained system.

These equations rarely have solutions in $\H$, and are, accordingly,
usually solved in some enlargement $\CV$ of $\H$ \ci{Hajicek}.  Since
the inner product of $\H$ is not defined on $\CV$, this leads to
certain problems.  Moreover, it is not {\em a priori} 
clear which enlargement to
choose. For example, when $\hat{\phv}_i$ is a second-order
differential operator, a case which is of prime importance for quantum
cosmology, there will be two linearly independent eigenfunctions
$\Ps_0^i$, $i=1,2$, with eigenvalue zero. In cosmological models,
where $\hat{\phv}_i$ is the Wheeler-DeWitt operator (that is, the
quantized super-Hamiltonian), the wave-function of the Universe is
supposed to be a certain linear combination of $\Ps_0^1$ and
$\Ps_0^2$.  The formalism doesn't tell which combination to choose.

Partly in response to these problems, and partly on the basis of
purely mathematical considerations, the author's method of constrained
quantization singles out the second step of the classical symplectic
reduction procedure as the one to be quantized.  We will first review
this method in its original formulation, and then introduce a
refinement. Upon the latter, our method turns out to be surprisingly
closely related to Dirac's method, applied to the enlargement $\CV$,
without sharing the difficulties just mentioned.  A related approach
is discussed by Marolf \ci{Marolf}.
\section{Quantized symplectic reduction}
Applied to the special case of Marsden-Weinstein reduction of
cotangent bundles, as above, our method amounts to choosing a dense
subspace $\CE$ of $L^2(Q)$ on which the quadratic form \be
(\Ps,\Ph)_0=\int_G dx\, (\Ps, U(x)\Ph)_{L^2(Q)} \ll{psph0} \ee is well
defined. Here $U$ is a unitary \rep\ of $G$ on $L^2(Q)$ which
quantizes the $G$-action on $Q$, and $dx$ is a Haar measure on $G$.
Unless $G$ is compact, this form tends to be unbounded and even
non-closable on $\CE$, but one can nonetheless form the quotient
$\CE/\CN$ of $\CE$ by the null space $\CN$ of $(\, ,\,)_0$. When the
latter is positive semi-definite, $\CE/\CN$ is a pre-Hilbert space in
the sesquilinear form inherited from $(\, ,\,)_0$. Its completion
$\H^0$ is the physical state space of the constrained system, and as
such serves as the quantization of classical reduced space
$J\inv(0)/G=T^*(Q/G)$.  This is true whether or not the
Marsden-Weinstein quotient is regular.

The classical physical observables are functions $F$ on $T^*Q$ which
Poisson-commute with the $G$-action.  The quantization $\hat{F}$ of
such an observable on the unphysical \Hs\ $L^2(Q)$ should commute with
$U(G)$, and leave $\CE$ stable. If these two conditions hold,
$\hat{F}$ leaves $\CN$ stable, and therefore quotients to an operator
$\hat{F}^0$ on $\CE/\CN$. If $\hat{F}$ is bounded, $\hat{F}^0$ is
bounded under suitable assumptions on $G$ \ci{MT}, and
may be extended to $\H^0$ by continuity. When $\hat{F}$ is unbounded,
containing $\CE$ in its domain, this procedure yields an unbounded
operator on the dense subspace $\CE/\CN$ of $\H^0$. The point is now
that in typical examples $\hat{F}^0$ tends to be essentially
self-adjoint on the projection $\CE^0$ of $\CE$ to $\H^0$. This is
possible, because $\CE^0$ is generally larger than $\cci(Q/G)$.

To see this in a simple example, classically due to Gotay and Bos
  \ci{GotayBos}, consider the standard action of $G=SO(2)$ on
  $Q=\R^2$.  The \mom\ for the pull-back action on $S=T^*\R^2$ is
  $J(p,q)=q\wedge p$, so that the quotient $J\inv(0)/SO(2)$ may be
  identified with $\R_0^+\x \R$ (where $\R_0^+:=\R^+\cup 0$), which
  may be thought of as the cotangent bundle $T^*\R_0^+$.  Hence the
  singularity in the reduced space takes the form of a boundary.

Quantum reduction is done with $\H=L^2(\R^2)$ and $\CE=\cci(\R^2)$.
 The unconstrained quantum Hamiltonian $\hat{H}_{\mbox{\tiny
 phys}}=-\Dl+V(r)$ on $L^2(\R^2)$ is $SO(2)$-invariant and essentially
 self-adjoint on $\cci(\R^2)$.  Using the unitary transformation
 $U:L^2(\R^+,rdr)\raw L^2(\R^+,dr)$ defined by
 $U\Ps(r):=\sqrt{r}\Ps(r)$, we have \be U\hat{H}^0_{\mbox{\tiny
 phys}}U^*=-\frac{d^2}{dr^2} -\frac{1}{4r^2} +V(r).  \ll{Upitiny} \ee
 While the analysis of this expression is quite straightforward for
 any reasonable potential $V$, the free case $V=0$ already suffices to
 illustrate the main point.

 Defined on $\cci(\R^+)$, the operator (\ref{Upitiny}) is in the limit
 circle case \ci{ReedSimon75} at 0 and in the limit point case
 \ci{ReedSimon75} at $\infty$. Hence it has deficiency indices
 $(1,1)$, so that it is not essentially self-adjoint. However, defined
 on $\CE^0=\cci(\R^2)^0$, which consists of functions of the type
 $\Ps(r)=\sqrt{r}f(r^2)$ with $f\in\cci(\R^+_0)$, the operator in
 question is essentially self-adjoint, as follows from Thm.\ 3 in
 Nussbaum \ci{Nussbaum}.  The closure of the latter operator is an
 extension of the closure of the former, to whose domain one adds
 functions of the indicated type in order to achieve essential
 self-adjointness. The boundary condition $\Ps(0)=0$ corresponds to a
 hard wall potential at the origin.

Other examples of this phenomenon are given by Wren \ci{Wren}, who
looks at the quantization of Stieffel chambers (i.e., quotients of a
maximal torus of a compact Lie group by its Weyl group), finding that
our procedure assigns Neumann boundary conditions to the Laplacian.
\section{Minisuperspace quantum cosmology}
Let us now apply our quantization method to the simplest cosmological
models in minisuperspace.  Using the canonical formalism (see our
earlier work \ci{WW} for a covariant approach), we assume that the
Hamiltonian constraint $H=0$, in which $H$ is a scalar, is the only
constraint on the classical minisuper phase space. If the \Hs\ $\H$ 
results from
the quantization of the unconstrained phase space, and the
Wheeler-DeWitt operator $\hat{H}$ is the quantization of the
Hamiltonian constraint (realized as an unbounded self-adjoint operator
on $\H$), our formalism stipulates that firstly one should find a
dense subspace $\CE\subset\H$ on which the quadratic form \be
(\Ps,\Ph)_0=\int_{\R}dt\,(\Ps,e^{itH}\Ph) \ll{ip0} \ee (cf.\
(\ref{psph0})) is well-defined. Secondly, one should determine the
reduced \Hs\ $\H^0$, defined as in the preceding section.  It would
seem, therefore, that one needs to explicitly compute the null space
$\CN$ of $(\, ,\,)_0$. Fortunately, in practice there is often no need
to do so. One may, instead, start from an {\em Ansatz} \Hs\
$\til{\H}^0$, and find a `quantum reduction map' $V:\CE\raw\til{\H}^0$
satisfying \be (V\Ps,V\Ph)_{\til{\H}^0}=(\Ps,\Ph)_0. \ll{V0} \ee It is
easily seen that $V$ quotients and extends to a unitary isomorphism
between $\H^0$ and $\til{\H}^0$, through which one may transfer the
physical Hamiltonian and other observables from the unknown space
$\H^0$ to $\til{\H}^0$, where everything is explicit \ci{MT}.  The
operators $\hat{A}$ on $\H$ that commute with $\hat{H}$ and leave
$\CE$ stable, are represented by physical observables $\hat{A}^0$ on
$\til{\H}^0$ by the prescription \be \hat{A}^0
V\Ps=V\hat{A}^0\Ps. \ll{obs} \ee

In the case at hand, these two steps may be taken at one stroke.
Using the theory of eigenfunction expansions \ci{Berezanski,PSW}, one
finds a Hilbert space $\H_-$ and a Hilbert-Schmidt injection
$\H\hookrightarrow\H_-$, such that (almost) all {\em relevant}
generalized eigenfunctions of $\hat{H}$ (that is, those contributing
to the spectral resolution of $\H$) lie in $\H_-$. Hence $\H$ is a
Hilbert subspace of $\H_-$ in the sense of Schwartz \ci{Schwartz}
(i.e., the embedding $\H\hookrightarrow\H_-$ is continuous). If $\H_+$
is the continuous dual of $\H_-$, one obtains an anti-linear map
$\Ps\raw\til{\Ps}$ from $\H_+$ to $\H$, defined by \be
\la\Ps,\Ph\ra=(\til{\Ps},\Ph)_{\H} \ll{sbc} \ee for all
$\Ph\in\H$. Here the restriction of $\Ps\in\H_+$ to $\H\subset\H_-$
(with its own \Hs\ topology) is continuous because of the continuity
of the embedding $\H\hookrightarrow\H_-$; the Riesz-Fischer theorem
then implies the existence of an element $\til{\Ps}\in\H$ for which
(\ref{sbc}) holds.  The image $\til{\H}_+$ of $\H_+$ in $\H$ under
this map is always dense, and when $\H$ is dense in $\H_-$, the map
$\Ps\raw\til{\Ps}$ is injective, so that $\H_+$ may be seen as a dense
Hilbert subspace of $\H$.  One then has a Gel'fand triplet (or rigged
\Hs) $\H_+\subset\H\subset\H_-$.  However, it is by no means necessary
that $\H$ be dense in $\H_-$, so that the formalism of Schwartz is a
generalization of that of Gel'fand.

In case that the spectrum $\sg(\hat{H})$ does not contain 0, the
physical \Hs\ $\H^0$ determined by (\ref{sbc}) is empty.  When
$\hat{H}$ has 0 in its discrete spectrum $\sg_d(\hat{\H})$, so that
the Wheeler-DeWitt equation $\hat{H}\Ps=0$ has a solution $\Ps\in\H$,
there is no need for our formalism. This rarely occurs in
minisuperspace models; instead, we assume that 0 lies in the
absolutely continuous spectrum $\sg_{ac}(\hat{\H})$.  We may take
$\til{\H}^0$ as the corresponding multiplicity space, realized in some
arbitrary fashion, with \be
V\til{\Ps}(\mu)=\overline{\la\Ps,\phv_{\mu}(0)\ra}, \ll{VPs} \ee
defined for $\Ps\in\H_+$, so that $\til{\Ps}\in\CE$.  Here
$\phv_{\mu}(\lm)$, where $\lm\in\sg(E)$ and $\mu$ labels a basis of
$\til{\H}^0$, is a generalized eigenfunction of $\hat{H}$ in $\H_-$
with generalized eigenvalue $\lm$. Note that (\ref{VPs}) is well
defined, for the $\phv_{\mu}(\lm)$ lie in the closure of $\H$ in
$\H_+$.  One easily verifies (\ref{V0}) from (\ref{ip0}) and eq.\
(1.17) in \S V.1 of Berezanskii's book \ci{Berezanski}.

An operator $\hat{A}$ on $\H$ that leaves $\H_+$ stable has a dual
$\hat{A}^*:\H_-\raw\H_-$. If, in addition, $\hat{A}$ commutes with
$\hat{H}$, the dual $\hat{A}^*$ cannot change the generalized
eigenvalue $\lm$, so that, in particular, one has
$\hat{A}^*\phv_{\mu}(0)=\hat{A}_{\mu\nu}^*\phv_{\nu}(0)$ for certain
coefficients $\hat{A}_{\mu\nu}^*$.  Combining this with (\ref{VPs})
and (\ref{obs}), we see that an observable $\hat{A}_{\mbox{\tiny
phys}}$ on $\H$ is realized on the physical state space $\til{\H}^0$
by \be \hat{A}^0_{\mbox{\tiny
phys}}f(\mu)=\overline{\hat{A}^*_{\mu\nu}}f(\nu). \ll{hatA} \ee
\section{The wave-function of the Universe}
Physical processes in the quantum Universe may be computed if one
specifies the physical Hamiltonian; as pointed out before, this should
be an operator $\hat{H}_{\mbox{\tiny phys}}$ on the unconstrained \Hs\
$\H$ which commutes with the Wheeler-DeWitt operator $\hat{H}$, and
leaves $\H_+$ stable.  This operator induces an operator
$\hat{H}^0_{\mbox{\tiny phys}}$ on the physical \Hs\ $\til{\H}^0$,
given by (\ref{hatA}).  One may anticipate that the spectrum of
$\hat{H}^0_{\mbox{\tiny phys}}$ is purely continuous, since otherwise
the Universe would settle in its ground state, and nothing would
happen.

For example, a classical homogeneous and isotropic universe filled with
noninteracting dust may be described by the phase space $T^*\R^2$,
with co-ordinates $(\al,\phi)$ standing for the logarithm of the
radius and the dust field, respectively, with conjugate momenta
$(p_{\al},p_{\phi})$. The classical super-Hamiltonian is \be
H_{\kp}=\half\left(p_{\al}^2+\kp e^{4\al}-p_{\phi}^2\right) , \ee
where $\kp=0$ or $\pm 1$. A reasonable choice for
the classical physical Hamiltonian is \be
H_{\mbox{\tiny phys}}= p_{\phi}; \ll{physH} \ee the idea is that the
dust field serves as a physical time parameter \ci{Isham}, but since
$\phi$ itself does not Poisson-commute with the constraint $H_{\kp}$,
this idea must be implemented through its conjugate $p_{\phi}$, which
does.

The unconstrained \Hs\ is simply $\H=L^2(\R^2)$, with wave-functions
$\Ps(\al,\phi)$. The Wheeler-DeWitt equation \be
\half\left(-(\partial/\partial\al)^2+\kp\exp(4\al)+ (\partial/\partial
\phi)^2\right)\Ps=0 \ee can be solved explicitly for all values of
$\kp$ \ci{WW}; we now add the knowledge that the {\em relevant}
generalized eigenfunctions must be polynomially bounded \ci{PSW} in
order to contribute to the spectral resolution of $\H$. For $\kp=0$
this leads to $\til{\H}^0_0=L^2(\R)\ot\C^2$, whereas for $\kp=\pm 1$
one has $\til{\H}^0_{\pm 1}=L^2(\R)$. The generalized eigenfunctions
at 0 are, omitting the argument (0), \bea
\phv^{\kp=0}_{k,\pm}(\al,\phi) & = & \exp[ik(\al\pm\phi)]; \\
\phv^{\kp=1}_{k}(\al,\phi) & = & \pi^{-1}e^{i\phi k}\sqrt{\sinh(\pi
|k|/2)}K_{i|k|/2}(\half e^{2\al});\\ \phv^{\kp=-1}_{k}(\al,\phi) & = &
\half e^{i\phi k}\sqrt{{\rm cosech}(\pi |k|/2)}
(J_{i|k|/2}+J_{-i|k|/2})(\half e^{2\al}).  \eea

With (\ref{physH}) as the physical Hamiltonian, it is reasonable to
ask the wave-function of the Universe to peak around $k=0$; this
approximately yields $\Ps^{\kp=0}\simeq 1$, $\Ps^{\kp=1}\simeq K_0$,
and $\Ps^{\kp=0}\simeq J_0$. This agrees with Hartle-Hawking 
boundary conditions for
$\kp=-1$ only, and always disagrees with Vilenkin boundary conditions 
(as imposed by Zhuk \ci{Zhuk}).  However, a direct
comparison is difficult, because these authors  use a
mathematical framework unrelated to the theory
of self-adjoint operators on a Hilbert space.
\section*{Acknowledgments}
The author is financially supported by a fellowship from the Royal
Netherlands Academy of Arts and Sciences (KNAW).  He is grateful to
Erik Thomas for introducing him to L. Schwartz' theory of Hilbert
subspaces.

\end{document}